\documentclass[conference]{IEEEtran}
\IEEEoverridecommandlockouts
\usepackage{cite}
\usepackage{amsmath,amssymb,amsfonts}
\usepackage{algorithmic}
\usepackage{graphicx}
\usepackage{textcomp}
\usepackage{xcolor}
\usepackage[thinlines]{easytable}
\usepackage{stackengine}
\usepackage{array}

\usepackage{ragged2e}
\usepackage{graphicx}
\usepackage[export]{adjustbox}

\usepackage{blindtext}

\usepackage{adjustbox}
\usepackage{multirow}

\usepackage{breakcites}
\usepackage{cite}
\usepackage{bm}

\usepackage{mathtools}

\usepackage[center]{caption}

\usepackage[edges]{forest}

\usepackage{float}

\usepackage{inconsolata}
 \usepackage{url}
\usetikzlibrary{fit}

\definecolor{burgundy}{rgb}{0.5, 0.0, 0.13}
\definecolor{brownweb}{rgb}{0.65, 0.16, 0.16}
\definecolor{chromeyellow}{rgb}{1.0, 0.65, 0.0}
\definecolor{babypink}{rgb}{0.96, 0.76, 0.76}
\definecolor{melon}{rgb}{0.99, 0.74, 0.71}
\definecolor{moccasin}{rgb}{0.98, 0.92, 0.84}

\newcommand\xrowht[2][0]{\addstackgap[.5\dimexpr#2\relax]{\vphantom{#1}}}
\def\BibTeX{{\rm B\kern-.05em{\sc i\kern-.025em b}\kern-.08em
    T\kern-.1667em\lower.7ex\hbox{E}\kern-.125emX}}
\begin{document}

\title{SpaML: a Bimodal Ensemble Learning Spam Detector based on NLP Techniques 
}

\author{\IEEEauthorblockN{
Jaouhar Fattahi}
\IEEEauthorblockA{{Department of Computer Science and Software Engineering}\\
{Laval University, }Quebec city, Canada.\\
jaouhar.fattahi.1@ulaval.ca}
\and
\IEEEauthorblockN{ Mohamed Mejri}
\IEEEauthorblockA{{Department of Computer Science and Software Engineering}\\
{Laval University, }Quebec city, Canada.\\
mohamed.mejri@ift.ulaval.ca}

}

\maketitle

\begin{abstract}
In this paper, we put forward a new tool, called SpaML, for spam detection using a set of supervised and unsupervised classifiers, and two techniques imbued with Natural Language Processing (NLP), namely Bag of Words (BoW) and Term Frequency-Inverse Document Frequency (TF-IDF). We first present the NLP techniques used. Then, we present our classifiers and their performance on each of these techniques. Then, we present our overall Ensemble Learning classifier and the strategy we are using to combine them. Finally, we present the interesting results shown by SpaML in terms of accuracy and precision.

\end{abstract}


\textit{This paper  was  accepted, on October 13, 2020, for publication and oral presentation at the 2021 IEEE 5th International Conference on Cryptography, Security and Privacy (CSP 2021) to be held in Zhuhai, China during January 8-10, 2021 and hosted by Beijing Normal University (Zhuhai).}

\section*{Notice}

\textbf{\copyright 2021 IEEE. \textit{Personal use of this material is permitted. Permission from IEEE must be obtained for all other uses, in any current or future media, including reprinting/republishing this material for advertising or promotional purposes, creating new collective works, for resale or redistribution to servers or lists, or reuse of any copyrighted component of this work in other works.}}

\section{Introduction}

Spam is mass text-based e-mails, such as advertising mailings, sent over the Internet. These texts are usually sent to thousands of e-mail addresses without being solicited.Apart from benign but annoying commercial spam, it often contains malicious hyperlinks pointing to various types of viruses, or phishing texts to lure individuals into providing sensitive data such as personal information, banking and credit card details, and passwords. This information is usually used to access important accounts and can lead to identity theft and financial loss. Spam can also be received on mobile phones in the form of short text messages (i.e sms). Most of them do not originate from another phone. Instead, they stem from a computer and are sent to one's phone, at virtually no cost to the sender, using an email address or an instant messaging account. Clicking a link in a spam text message can set up malicious software that can gather information about one's phone. This can also affect the performance of one's mobile phone by eating away at its memory space. This can also result in undesirable charges on one's mobile phone invoice. Identifying a message as spam is not an easy thing. Spam filters are usually offered by email or mobile phone providers. Many of these programs automatically check the contents of emails and blacklist the senders. This process works via a word list with typical spam phrases and expressions. However, no method can guarantee that spam is automatically and systematically recognized. In addition, it also happens that some important and expected emails end up in junk mail because of these filters. In this paper, we address this problem from a machine learning (ML) perspective. We propose a tool, which we call SpaML, for spam detection. This tool is based on a set of supervised and unsupervised machine learning models and relies on two Natural Language Processing (NLP) techniques that make up its two modes.


\section{Paper organization}

The rest of this paper is organized as follows. In Section \ref{sec3}, we give an overview of the Natural Language Processing (NLP) techniques we are using to convert messages into vectors of numbers. In Section \ref{sec5}, we present the SpaML architecture, as well as the base detectors it exploits.  In Section \ref{sec6}, we present our experiments, as well as the interesting results shown by SpaML regarding its accuracy and precision.  In Section \ref{sec7}, we discuss the results and we compare our research to other related pieces of research addressing the same problem.  In Section \ref{sec8}, we draw conclusions.

\section{Natural Language Processing Techniques} \label{sec3}

Natural Language Processing (NLP) is a field of Artificial Intelligence (AI). It has been used to predict diseases\cite{WashingtonGMAGWHL08}, analyze sentiment\cite{colingMurrayCJ18}, identify fake news\cite{icaartZhouGBH19}, recruit talent\cite{LytvynVR19}, detect cyberterrorism exchanges \cite{Castillo-Zuniga20}, etc. In this paper, we focus on two NLP techniques that convert natural texts into a vectors of numbers. These vectors will become the inputs of our ML models, which are used by our tool SpaML, to predict whether a vector, thus a text, is spam or ham (ham is another word for regular in this context). These two techniques are Bag of Words (BoW) and Term Frequency-Inverse Document Frequency (TF-IDF).

\subsection{Bag of words}

Bag of words  is the most basic technique of converting a text into a vector of numbers. The following is an example of how this works. Let us consider these three documents:

\begin{center}
\begin{tabular}{ l l }
$d_1$: & this is a dog \\ 
$d_2$: &  this is not a dog \\  
$d_3$: & a dog is a special pet which is a friendly pet
\end{tabular}
\end{center}

First, we build a lexicon (i.e. set of vocabulary) from all the words appearing in all documents. The lexicon here is \{\textit{this, is, a, dog, not, special, pet, which, friendly}\}. Then, we take each of these words and score its occurrence in the text with 1 when the word exists and with 0 when the word does not. Then we sum the occurrences up. This is shown in Table \ref{tab:mytable1}.

\begin{center}
 \begin{table}[h]
\centering
    \caption{Bag of words}
    \label{tab:mytable1}
\scalebox{0.78}{
\begin{tabular}{| c| c|c| c| c|c|c|c|c|c|}
 \hline\xrowht[()]{10pt}
   \textbf{Times in}           &   \textbf{this}&   \textbf{is} &  \textbf{a}&  \textbf{dog} & \textbf{not}&  \textbf{special}&  \textbf{pet}  &  \textbf{which} &  \textbf{friendly}\\ 
 \hline\xrowht[()]{10pt}
$d_1$ & 1 &  1&  1&  1&  0& 0&0& 0& 0\\ 
 \hline\xrowht[()]{10pt}
$d_2$&   1&  1&  1&  1& 1 &0 &0&0&0\\  
 \hline\xrowht[()]{10pt}
$d_3$&  0&  2& 3 & 1&  0& 1&2&1&1\\
\hline
\end{tabular}
}
 \end{table}
\end{center}

The resulting vectors for the documents $d_1$, $d_2$, and $d_3$ are (1,1,1,1,0,0,0,0,0), (1,1,1,1,1,0,0,0,0), (0,2,3,1,0,1,2,1,1), respectively.

This is the main idea behind BoWs. However, in practice, with long texts, we do not consider all the words in all documents to build the vectors of numbers. Instead, we only consider the most frequent words appearing in the documents by adding their occurrences in each document and ordering them. The number of the most frequent words tightly depends on its relevance to the study goals.

\subsection{Term Frequency-Inverse Document Frequency }

Term Frequency-Inverse Document Frequency is a more elaborated way to represent texts into vectors of numbers. In this technique, we first calculate the term frequency as follows:

\begin{equation} \text{TF}_{t,d} = \frac{\text{N}_{t,d}}{\text{NT}_d}  \label{eq1}\end{equation}

where $\text{N}_{t,d}$ is the number of times the term $t$ shows up in the document $d$, and ${\text{NT}_d}$ is the number of terms in the document $d$. Hence, every document has its own term frequency.

Then, we calculate the Inverse Document Frequency (IDF) for every term as follows:

\begin{equation} \text{IDF}_t = \text{Log (} \frac{{\text{ND}}}{\text{ND}_t}\text{)}  \label{eq2}\end{equation}

where $\text{Log}$ is the Logarithmic function, $\text{ND}$ is the number of documents, and $\text{ND}_t$ is the number of documents with the term $t$. The IDF measures the importance of a term for the documents. Finally, we calculate the  Term Frequency-Inverse Document Frequency (TF-IDF) score for every term in the documents as follows:

$$\text{TF}\_\text{DF}_{t,d} = \text{TF}_{t,d} \times \text{IDF}_t$$

The higher is the $\text{TF}\_\text{DF}_{t,d} $, the more important is the term. To be concrete, let us take the example of the previous subsection.  Since $d_1$ contains $4$ terms and the term  \textit{this} appears just once in $d_1$, then $\text{TF}_{d_1, this} = \frac{1}{4}$. The same thing for $\text{TF}_{d_1, is}$, $\text{TF}_{d_1, a}$, and $\text{TF}_{d_1, dog}$. Since the term \textit{not} does not appear in $d_1$, then $\text{TF}_{d_1, not} = 0$. The same thing for $\text{TF}_{d_1, special}$, $\text{TF}_{d_1, pet}$, $\text{TF}_{d_1, which}$, and $\text{TF}_{d_1, friendly}$.  Table \ref{tab:mytable2} presents the TF values for all the terms in the three documents.


\begin{center}
    \begin{table}[h]
\centering
    \caption{Term Frequency}
    \label{tab:mytable2}
\scalebox{0.78}{

\begin{tabular}{| l| l| l| l| l| l| l| }
 \hline\xrowht[()]{10pt}
      \textbf{Term}        &  N$_{d_1}$&  N$_{d_2}$ & N$_{d_3}$& TF$_{d_1}$ & TF$_{d_2}$ & TF$_{d_3}$ \\ 
 \hline\xrowht[()]{10pt}

      \textbf{this}        &  $1$&  $1$ & $0$& $\frac{1}{4}$ & $\frac{1}{5}$ & $0$ \\ 
 \hline\xrowht[()]{10pt}

      \textbf{is}        &  $1$&  $1$ & $2$& $\frac{1}{4}$ & $\frac{1}{5}$ & $\frac{2}{11}$ \\ 
\hline\xrowht[()]{10pt}

      \textbf{a}        &  $1$&  $1$ & $3$& $\frac{1}{4}$ & $\frac{1}{5}$ & $\frac{3}{11}$ \\ 
 \hline\xrowht[()]{10pt}
      \textbf{dog}        &  $1$&  $1$ & $1$& $\frac{1}{4}$ & $\frac{1}{5}$ & $\frac{1}{11}$ \\ 
 \hline\xrowht[()]{10pt}
      \textbf{not}        &  $0$&  $1$ & $0$& $0$ & $\frac{1}{5}$ & $0$ \\ 
 \hline\xrowht[()]{10pt}
      \textbf{special}        &  $0$&  $0$ & $1$& $0$ & $0$ & $\frac{1}{11}$ \\ 
 \hline\xrowht[()]{10pt}
     \textbf{pet}        &  $0$&  $0$ & $2$& $0$ & $0$ & $\frac{2}{11}$ \\ 
 \hline\xrowht[()]{10pt}
      \textbf{which}        &  $0$&  $0$ & $1$& $0$ & $0$ & $\frac{1}{11}$ \\ 
 \hline\xrowht[()]{10pt}
      \textbf{friendly}        &  $0$&  $0$ & $1$& $0$ & $0$ & $\frac{1}{11}$ \\ 
 \hline
\end{tabular}
}

\end{table}
\end{center}

As for the IDF values, since we have three documents, NT$_d = 3$. Since the term \textit{this} appears in just two documents, $d_1$ and $d_2$, then IDF$_{this} = \text{Log}(\frac{3}{2})=0.176$. Table \ref{tab:mytable3} contains the IDF values for all terms. Once done, the calculation of the TF\_DF becomes straightforward. This is given in Table \ref{tab:mytable4}. The resulting vectors for the documents $d_1$, $d_2$, and $d_3$ are (0.036,0,0,0,0,0,0,0,0), (0.035,0,0,0,0.095,0,0,0,0), (0,0,0,0,0,0.043,0.087,0.043,0.043), respectively.

\begin{center}
    \begin{table}[h]
\centering
    \caption{Inverse Document Frequency }
    \label{tab:mytable3}
\scalebox{0.78}{

\begin{tabular}{| l| l| l| l| l|}
 \hline\xrowht[()]{10pt}
     \textbf{Term}        &  N$_{d_1}$&  N$_{d_2}$ & N$_{d_3}$& IDF   \\ 
 \hline\xrowht[()]{10pt}

     \textbf{this}        &  $1$&  $1$ & $0$& Log($\frac{3}{2}$) = $0.176$  \\ 
 \hline\xrowht[()]{10pt}

     \textbf{is}        &  $1$&  $1$ & $2$& Log($\frac{3}{3}$) = $0$  \\ 
\hline\xrowht[()]{10pt}

     \textbf{a}        &  $1$&  $1$ & $3$& Log($\frac{3}{3}$) = $0$  \\ 
 \hline\xrowht[()]{10pt}
     \textbf{dog}        &  $1$&  $1$ & $1$& Log($\frac{3}{3}$) = $0$  \\ 
 \hline\xrowht[()]{10pt}
     \textbf{not}        &  $0$&  $1$ & $0$& Log($\frac{3}{1}$) = $0.477$  \\ 
 \hline\xrowht[()]{10pt}
     \textbf{special}        &  $0$&  $0$ & $1$& Log($\frac{3}{1}$) = $0.477$ \\ 
 \hline\xrowht[()]{10pt}
     \textbf{pet}        &  $0$&  $0$ & $2$& Log($\frac{3}{1}$) = $0.477$ \\ 
 \hline\xrowht[()]{10pt}
     \textbf{which}        &  $0$&  $0$ & $1$& Log($\frac{3}{1}$) = $0.477$  \\ 
 \hline\xrowht[()]{10pt}
     \textbf{friendly}        &  $0$&  $0$ & $1$& Log($\frac{3}{1}$) = $0.477$  \\ 
 \hline
\end{tabular}
}

\end{table}
\end{center}


\begin{center}
    \begin{table}[h]
\centering
    \caption{Term Frequency-Inverse Document Frequency}
    \label{tab:mytable4}
\scalebox{0.8}{

\begin{tabular}{| l| l| l| l| }
 \hline\xrowht[()]{10pt}
     \textbf{Term}        &  \textbf{TF\_IDF$_{d_1}$} & TF\_IDF$_{d_2}$ & TF\_IDF$_{d_3}$ \\ 
 \hline\xrowht[()]{10pt}

     \textbf{this}        &   $\frac{1}{4}*0.176 = 0.036$ & $\frac{1}{5}*0.176= 0.035$ & $0*0.176=0$ \\ 
 \hline\xrowht[()]{10pt}

     \textbf{is}        &   $\frac{1}{4}*0 = 0$ & $\frac{1}{5}*0 = 0$ & $\frac{2}{11}*0 = 0$ \\ 
\hline\xrowht[()]{10pt}

     \textbf{a}        &   $\frac{1}{4}*0 = 0$ & $\frac{1}{5}*0 = 0$ & $\frac{3}{11}*0 = 0$ \\ 
 \hline\xrowht[()]{10pt}
     \textbf{dog}        & $\frac{1}{4}*0 = 0$ & $\frac{1}{5}*0 = 0$ & $\frac{1}{11}*0 = 0$ \\ 
 \hline\xrowht[()]{10pt}
     \textbf{not}         &0 * 0.477 = 0& $\frac{1}{5}* 0.477 = 0.095$ & $0* 0.477 = 0$ \\ 
 \hline\xrowht[()]{10pt}
     \textbf{special}        & $0* 0.477 = 0$ & $0* 0.477 = 0$ & $\frac{1}{11}* 0.477 = 0.043$ \\ 
 \hline\xrowht[()]{10pt}
     \textbf{pet}       & $0* 0.477 = 0$ & $0* 0.477 = 0$ & $\frac{2}{11}* 0.477 = 0.087$ \\ 
 \hline\xrowht[()]{10pt}
     \textbf{which}        & $0* 0.477 = 0$ & $0* 0.477 = 0$ & $\frac{1}{11}* 0.477 = 0.043$ \\ 
 \hline\xrowht[()]{10pt}
     \textbf{friendly}       & $0* 0.477 = 0$ & $0* 0.477 = 0$ & $\frac{1}{11}* 0.477 = 0.043$ \\ 
 \hline
\end{tabular}
}

\end{table}
\end{center}

\section{SpaML architecture} \label{sec5}

\begin{figure}[hbt!]
  \centering

\scalebox{0.75}{
\forestset{
 default preamble={
 font=\Large,
 }
}
\begin{forest}
  for tree={
    draw,
    align=center,
    font=\footnotesize
  },
  forked edges,
  [Overall prediction \\ (Majority of votes), fill=melon 
    [ SpaML, fill=moccasin, circle,draw
      [ Prediction, fill=melon
          [MNB , fill=moccasin
          ]
      ]
      [ Prediction , fill=melon
          [LR , fill=moccasin
          ]
      ]
      [ Prediction, fill=melon
          [SVM , fill=moccasin
          ]
      ]
      [ Prediction, fill=melon
          [NCC , fill=moccasin
          ]
      ]
      [ Prediction, fill=melon
          [Xgboost , fill=moccasin
          ]
      ]
      [ Prediction, fill=melon
          [KNN , fill=moccasin
          ]
      ]
      [ Prediction, fill=melon
          [Perceptron , fill=moccasin
          ]
      ]
    ]
  ]
\end{forest}
}
  \caption{SpaML architecture}\label{fig:spamlArch}
\end{figure}
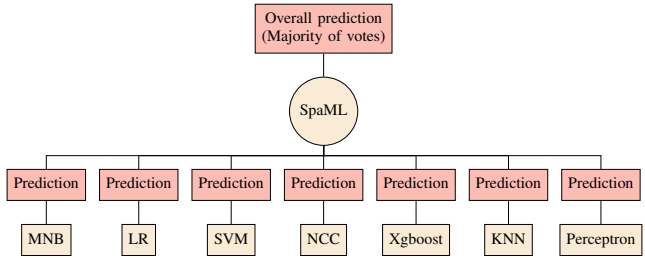

SpaML is a super learner whose architecture is given by Fig \ref{fig:spamlArch}. It is bimodal. That is to say, it can operate in two modes: BoW or TF-IDF, depending on the NLP technique selected by the user. It utilizes seven supervised and unsupervised detectors, namely MNB, LR, SVM, NCC, Xgboost, KNN and Perceptron based on Multinomial Naive Bayes, logistic regression, support sector machine, nearest centroid,  Extreme Gradient Boosting, K-nearest neighbors, and Perceptron algorithms, respectively. It uses the majority of vote strategy to make the final decision founded on the prediction of its base learners. That is to say, if three base learners vote spam (i.e. 1), and four vote ham (i.e. 0), then the result is ham, for instance. Below is a reminder of the basics of our base classifiers and how they work. 

\subsection{Multinomial Naive Bayes }

The Naive Bayes classifier (NBC)\cite{Webb2010459862} is a probabilistic and supervised model based on Bayes' theorem\cite{sep-bayes-theorem}. The theorem assumes that the features are all independent. That is why the algorithm is referred to as naive. To understand how the Naive Bayes classifier derives from Bayes theorem, assume a feature vector $X=(x_1, x_2, \mbox{...}, x_n)$ and a class variable $y_k$ among $K$ classes in the training data. Bayes's theorem implies: 

\begin{equation} P(y_k|X)=\frac{P(X|y_k).P(y_k)}{P(X)}, k \in \{1,2,\mbox{...}, K\}  \label{naive_eq1}\end{equation}

Considering the chain rule \cite{wiki:chain_rule} for multiple events:
\begin{equation} P(A_1  \cap A_{2} \cap \mbox{ ... } \cap A_n)= P(A_1 | A_{2} \cap \mbox{...}\cap A_n). P(A_{2} \cap \mbox{...} \cap A_n)  \label{naive_eq2}\end{equation}

The likelihood $P(X|y_k)$ can be written as follows:
\begin{equation}  \label{naive_eq3}
\begin{split}
P(X|y_k) & = P(x_1, x_2, \mbox{ ... }, x_n|y_k) \\
 & = P(x_1|x_2, \mbox{ ... }, x_n|y_k).P(x_2|x_3 \mbox{...}, x_n|y_k) \mbox{...}P(x_n|y_k)
\end{split}
\end{equation}

This is when the independence assumption of Bayes' theorem is useful, implying:

\begin{equation} P(x_i|x_{i+1} \mbox{...} x_n| y_k)=P(x_i|y_k), i \in \{1 \mbox{...} n\}  \label{naive_eq4}\end{equation}

The likelihood can so be reduced to:

\begin{equation}P(X|y_k) = \prod_{i=1}^{n} P(x_i|y_k) \label{naive_eq5}\end{equation}

The posterior probability $P(y_k|X)$ can also be reduced to: 
\begin{equation} P(y_k|X)=\frac{P(y_k).\prod_{i=1}^{n} P(x_i|y_k)}{P(X)},  k \in \{1,2, \mbox{...}, K\}  \label{naive_eq6}\end{equation}

Considering that $P(X)$ is a constant  for all $k \in \{1,2,\mbox{...}, K\}$, the Naive Bayes classification problem comes down to maximizing 

\begin{equation} P(y_k).\prod_{i=1}^{n} P(x_i|y_k)\label{naive_eq7}\end{equation}

The probability $P(y_k)$ is the relative frequency of the class $y_k$ in the training data. $P(x_i|y_k)$ can be calculated using usual distributions. The classifier is referred to as Multinomial Naive base (MNB) in case of multinomial distribution\cite{wiki:Multinomialdistribution} (in our case for two classes).


\subsection{Logistic regression}

Logistic regression (LR)\cite{LR123789} is a supervised model that uses a transformation called \textit{Logit} that calculates the logarithm of the probability of an event (e.g. a message being spam) divided by the probability of no event. It is defined as follows: 
\begin{center}
$\text{Logit}(p) = \text{Log}\displaystyle \frac{p}{1-p}$
\end{center}
where $p = p(y=\text{spam}|X)$ is the conditional probability of the output $y$ being spam knowing the input $X=(x_1, x_2, ..., x_n)$. 
Under the assumption of a linear relationship between $\text{logit}(p)$ and the predictors, ( i.e. $\text{Logit}(p)= \beta_0 + \beta_1 x_1 + ... + \beta_n x_n$), we have:


\begin{equation}  \label{log_eq1}
\displaystyle \frac{p}{1-p} =  \displaystyle {e^{\beta_0 + \beta_1 x_1 + ... + \beta_n x_n}}
\end{equation}


\begin{center}  \text{or else}\end{center} 

\begin{equation}  \label{log_eq2}
\displaystyle p = \displaystyle \frac{1}{1 + e^{-(\beta_0 + \beta_1 x_1 + ... + \beta_n x_n)}}
\end{equation}


As we can see it, $p$ is the sigmoid function applied to the weighted inputs. If it is close to $1$, then the event is present. If it is close to $0$, then it is not. 

\subsection{Support vector machine}

\begin{figure}[htbp]
\begin{minipage}{8 cm}
\includegraphics[scale=0.55]{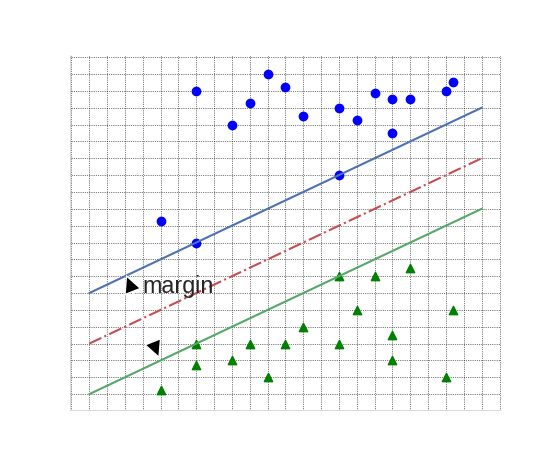}
\caption{SVM}
\label{fig:SVM1}
\end{minipage}
\end{figure}

A Support Vector Machine (SVM) is a supervised ML algorithm used for classification, regression, and anomaly detection. It is known for its strong theoretical basis \cite{SVMcortes1995support, SVM1481236}.  Its goal is to separate data into classes using a separation boundary so that the distance between the different classes of data and the boundary is as large as possible. This distance is also called \textit{margin} and the SVM is called \textit{wide-margin separator}. The \textit{support vectors} are generated by the data points closest to the boundary. They play a very important role in the model formation, and if they change, the position of the boundary will very likely change.  In Fig. \ref{fig:SVM1}, in two-dimensional space, the boundary is most likely the red line, the support vectors are most likely determined by the two points on the green line, as well as the two points on the blue line, and the margin is the distance between the boundary and the blue and red lines. Maximizing the margin reinforce noise-resistance and enables the model to be more generalizable. 

\subsection{Nearest Centroid}

The Nearest Centroid Classifier (NCC) belongs to the family of unsupervised ML algorithms. It classifies a data point in the class whose mean is closest to it (centroid is another word for mean). For a binary classification, the number of centroids is two. Initially, centroids are selected either manually, or randomly, or with the help of certain tools such as \textit{k-means++} in SKLearn. Then, each observation is assigned to the cluster whose centroid has the least squared Euclidean distance. Then, the new centroids are updated according to the observations in the modified clusters. The algorithm loops through the last two steps until the assignments no longer change. A point $P$ is assigned the class of the nearst centroid.

\subsection{Extreme Gradient Boosting}

Extreme Gradient Boosting   (Xgboost)\cite{XGBoostChenG16} is a supervised algorithm. It is basically a tree ensemble model consisting of a set of classification and regression trees where a score is associated with each leaf of the tree. The algorithm uses an objective function with a regularization term and optimizes it using the second order derivative as an approximation to score gains and make the best split and prune the trees. Xgboost is able to efficiently perform parallel processing which accelerates training, handle missing values, and make non-greedy tree pruning. Dropout and regularization mechanisms of Xgboost are particularly effective to reduce overfitting. 

\subsection{K-Nearest Neighbors}

The K-Nearest Neighbors (KNN, K $\in \mathbb{N}$) \cite{knn1053964}  is a very simple supervised classification algorithm. Its purpose is to classify target points of unknown classes according to their distance from the K points of a learning sample whose classes are known in advance. Each  point $P$ of the sample is considered as a vector of $\bm{\mathbb{R}}^n$, described by its coordinates $(p_1,p_2, ..., p_n)$. In order to determine the class of a target point $Q$, each of the K points closest to it takes a vote. The class of $Q$ is then the class with the majority of votes. KNN can use multiple types of distances \cite{DBLP:journals/corr/abs-1812-05944, thesis921292} in a normalized vector space to find the closest points to $Q$, such as the Euclidean distance, the Manhattan distance,  the Minkowski distance,  the Tchebychev distance, and the Canberra distance. The choice of the K parameter plays a crucial role in the accuracy and performance of the model.

\subsection{Perceptron}

Perceptron \cite{perceptron80230} is a supervised ML algorithm mainly used for binary classification. It uses a binary function that can decide whether an input, represented by a vector of numbers, belongs to a certain class or not. For an input vector $x=(x_1,...,x_n)$, $f$ is defined as follows: 
$$
f(x) = \left\{
    \begin{array}{ll}
        1 & \mbox{if } w.x + b > 0\\
        o & \mbox{if not.}
    \end{array}
\right.
$$

where $w$ is weight vector of size $n$, "." is the dot product operator, and $b$ is a bias. The algorithm starts by initializing the weights and the threshold $\gamma$. At time $t$, for a sample $j$ (i.e. $x^j$ and output $y^j$) of the training dataset $D$ of cardinality $s$, perform the two following steps:

\begin{enumerate}
\item compute the predicted candidate: $\hat{y}^j(t)= f(w(t) . x^j + b)$
\item update $w$:  $w_i(t+1) = w(t) + \rho . (y^j(t) - \hat{y}^j(t)) x^j_i$, for all features $i\in \{1,...,n\}$; $\rho$ is the learning rate.
\end{enumerate}

These two steps are repeated until the mean squared error $\frac{1}{s}\sum_{j=1}^{s} (y^j(t) - \hat{y}^j(t))^2$ becomes inferior to the threshold $\gamma$, or a predefined looping limit number is reached.

\section{Experiments and results} \label{sec6}

\subsection{Dataset}

We use the SMS dataset \cite{datset2034742} in our experiments. It consists of 5574 instances. 747 instances are spam messages, and 4827 are ham messages. To train our models, we randomly split the dataset into two parts: a training dataset and a test dataset. The first one contains 75\% of the original dataset items and the second one contains 25\% of them. Since the SMS dataset is imbalanced, we use  the \textit{stratify} strategy so that the selected records into the two resulting datasets keep the same distribution as the original one.

\subsection{Preprocessing}

Before converting texts into vectors of numbers we proceed to preprocessing it. First, we lowercase all texts. Then, we get rid of stop words, which are useless data such as \textit{the, a, an, in, for, because, that, over, more, him, each, who, these, into, below, are, by, etc}. Then, we perform a text stemming which is the process of reducing inflection in words to their root forms. For example, the words \textit{detection, detected, detecting, detector} are reduced to their stemming root word  \textit{detect}.


\subsection{Results}

To evaluate our classifiers, we use the two following scoring metrics:

\begin{center}
\text{Accuracy} $= \displaystyle \frac{\text{TP+TN}}{\text{TP + TN + FP + FN}}$
\end{center}

and,

\begin{center}
\text{Precision} $=\displaystyle \frac{\text{TP}}{\text{TP + FP}}$
\end{center}

where \text{TP} represents the true positives, \text{TN} the true negatives,  \text{FP} the false negatives, and \text{FP} the false positives. 

Accuracy is the fraction of labels that the model has managed to predict correctly. Precision is a good metric when false positives count. In spam detection, a false positive means that a message which is not actually spam has been identified as spam. In such a case, the user may lose important messages if the precision of the detection model is not high.

 Table \ref{tab:mytable5} summarizes the performance of the base detectors, as well as the overall detector, in both modes BoW and TF-IDF.

 \begin{table}[h]
\centering
    \caption{Classifier performance}
    \label{tab:mytable5}
\scalebox{0.8}{
\begin{tabular}{| l| c| c|c|c|}
 \hline\xrowht[()]{10pt}
  \multirow{2}{*}{\textbf{Classifier}} & \multicolumn{2}{|c|}{\textbf{BoW Mode}}  &  \multicolumn{2}{c|}{\textbf{\textbf{TF-IDF Mode}}} \\  \cline{2-5}\xrowht[()]{12pt}
                                                                    &\textbf{Accuracy (\%)} & \textbf{Precision} (\%)& \textbf{Accuracy (\%)} & \textbf{Precision} (\%)\\  \hline\xrowht[()]{10pt}
\textbf{MNB} & 96.04&96.16                                      & 96.08 &96.24\\  \hline\xrowht[()]{10pt}
\textbf{LR}            & 96.87&96.79                                      & 94.90 &96.89\\  \hline\xrowht[()]{10pt}
\textbf{SVM}                                        & 96.74&96.33                                      & 96.46 &96.45\\  \hline\xrowht[()]{10pt}
\textbf{NCC}               & 96.10&95.90                                      & 95.98 &95.56\\  \hline\xrowht[()]{10pt}
\textbf{Xgboost}                                & 96.65&96.24                                      & 96.13 &96.26\\  \hline\xrowht[()]{10pt}
\textbf{KNN}                                       & 90.93&95.27                                      & 89.76 &94.83\\  \hline\xrowht[()]{10pt}
\textbf{Perceptron}                          & 96.97&95.07                                      & 96.58 &94.94\\  \hline\xrowht[()]{10pt}

\textbf{SpaML}                                   & \textbf{98.11}&\textbf{98.91}    & \textbf{97.99} &\textbf{98.87}\\  
\hline
\end{tabular}
}
 \end{table}







\begin{figure}[htbp]
\centering
\begin{minipage}{7.5cm}
\includegraphics[width=7.5cm,height=4.5cm]{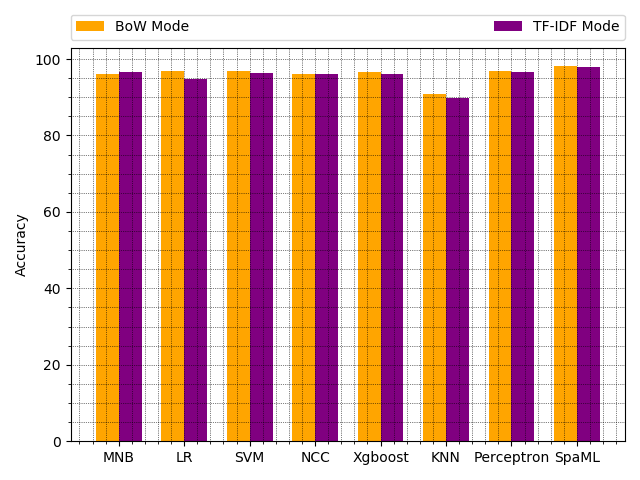}
\caption{Accuracy comparison (by Mode)}
\label{fig7}
\end{minipage}
\end{figure}

\begin{figure}[htbp]
\centering
\begin{minipage}{7.5cm}
\includegraphics[width=7.5cm,height=4.5cm]{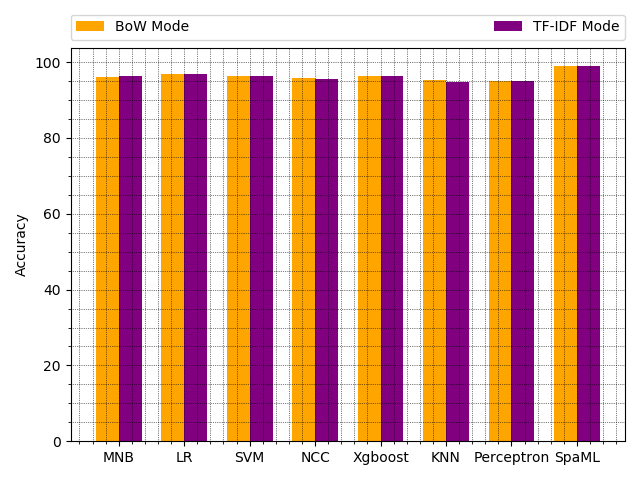}
\caption{Precision comparison (by Mode)}
\label{fig8}
\end{minipage}
\end{figure}

 Fig \ref{fig7} compares the accuracy of the models by mode and Fig \ref{fig8} compares the precision of the models by mode, as well.  

\begin{figure}[htbp]
\centering
\begin{minipage}{7cm}
\includegraphics[width=7cm,height=3cm]{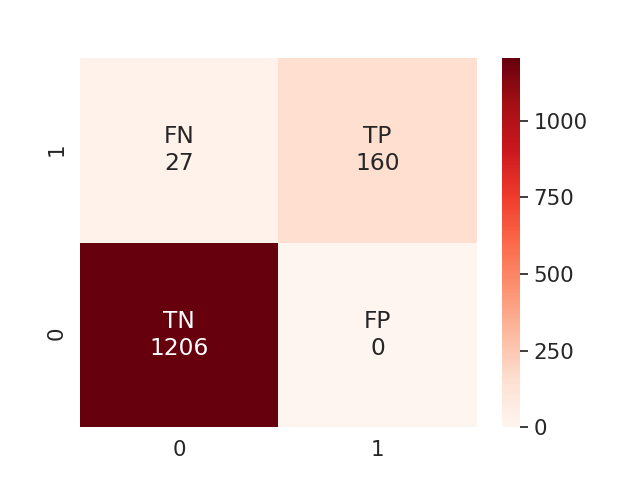}
\caption{SpaML confusion matrix (BoW mode)}
\label{fig9}
\end{minipage}
\end{figure}

\begin{figure}[htbp]
\centering
\begin{minipage}{7cm}
\includegraphics[width=7cm,height=3cm]{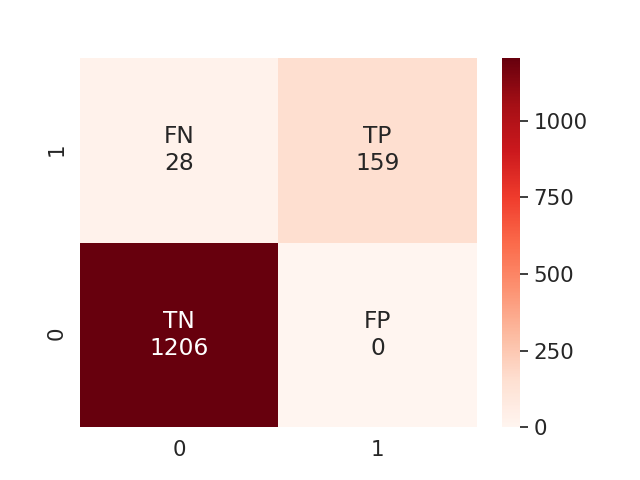}
\caption{SpaML confusion matrix (TF-IDF mode)}
\label{fig10}
\end{minipage}
\end{figure}

Fig \ref{fig9} shows the confusion matrix of SpaML in BoW mode on a sample of unseen data of 1393 records with 1206 ham messages and 187 spam messages. Fig \ref{fig10} shows the confusion matrix of SpaML in mode TF-IDF for the same records. It is worth mentioning that we have used a 10-fold cross-validation procedure to evaluate SpaML, as well as its base detectors, to make sure that they do not overfit data.

\section{Discussion and comparison with related work} \label{sec7}

In summary, in BoW mode, SpaML has displayed an accuracy of 98.11\% and a precision of 98.91\%. In TF-IDF mode, it has displayed an accuracy of 97.99\% and a precision of 98.87\%. This shows that SpaML copes very well with both modes. This being said, the difference between the two modes, although not very significant, gives a slight advantage to the BoW technique on the used dataset. Nevertheless, such a result cannot be generalized prior to evaluating SpaML on other datasets.  Other approaches like spam filtering based on adaptive statistical data compression models\cite{BratkoCFLZ06} using character-level or binary sequences have been proposed. They usually use dynamic Markov compression\cite{ItakuraC09} and partial matching\cite{NetoBC18} to evaluate the model. Rule-based filtering systems\cite{Xia20} using behavioral methods or linguistic methods have also been proposed to score and classify texts. Deep Learning\cite{DBLPMakkarK20, DBLPRoySB20, SaidaniAA19} is also beginning to be used to detect spam and algorithms such as CNN\cite{DBLPLiuL20g} and LSTM\cite{DBLPEryilmazSK20} are gaining ground. Hidden Markov Models \cite{DBLPGordilloC07, Dang7578397} have also been considered to address this problem. All these approaches and techniques give relatively good results. In our vision, these comparable  methods are not antagonistic or in competition with each other, on the contrary, they can be used in a collaborative context.

\section{Conclusion} \label{sec8}

In this paper, we have proposed a spam detector using two NLP-based techniques for text vectorization and a set of different classifiers, supervised and unsupervised. The super learner on top of these base learners, SpaML, has shown very interesting results in terms of precision and accuracy with the two techniques. This motivates us to explore other NLP techniques, as well as similar ones, to tackle this difficult problem of spam detection and extend it to other problems close to it,  such as tracking terrorism-related exchanges and targeting organized crime in social networks.

\section*{Acknowledgment}

This research was funded by the Natural Sciences and Engineering Research Council of Canada (NSERC).


\bibliography{biblio}

\bibliographystyle{IEEEtran}

\end{document}